\documentclass[journal]{IEEEtran}
\ifCLASSINFOpdf

\else
\fi
\usepackage{graphicx,amsmath,amssymb,cite}
\usepackage{color}
\usepackage{amsfonts}
\usepackage{amsmath}
\usepackage{graphicx}
\usepackage{amssymb}
\usepackage{stfloats}
\usepackage{subfig}
\usepackage{mathrsfs}
\usepackage{algorithm}
\usepackage{algorithmic}
\usepackage{booktabs}
\usepackage[numbers,sort&compress]{natbib}
\usepackage{comment}

\usepackage{etoolbox}
\pretocmd{\maketitle}{%
  \markboth{The definitive version was published in IEEE Network, vol. 36, no. 5, pp. 114-120, September/October 2022, doi: 10.1109/MNET.001.2200212.}{}%
}{}{}

\begin{document}

\title{Providing Location Information at Edge Networks: A Federated Learning-Based  Approach}

\author{Xin Cheng,~Tingting Liu,~Feng Shu,~Chuan Ma,~Jun Li,\\and~Jiangzhou Wang,~\IEEEmembership{Fellow,~IEEE}
\thanks{\emph{This work was supported in part by the National Natural Science Foundation of China (Nos. 62071234, 62071289, 61972093, 62171217 and under Grant No. 62002170), the Hainan Province Science and Technology Special Fund (ZDKJ2021022), the Scientific Research Fund Project of Hainan University under Grant KYQD(ZR)-21008, the National Key R\&D Program of China under Grant 2018YFB180110 and the China Postdoctoral Science Foundation under Grant No. 2021M691540. (Corresponding authors: Feng Shu and Chuan Ma).}}

\thanks{\emph{
Xin Cheng, and Jun Li are with School of Electronic and
Optical Engineering, Nanjing University of Science and Technology, Nanjing,
210094, China (e-mail:xincstar23@163.com).}}

\thanks{\emph{
Tingting Liu is with the Nanjing Institute of Technology, Nanjing 211167, China
and also with the Nanjing University of Science and Technology, Nanjing
210094, China.}}

\thanks{\emph{
Feng Shu is with the School of Information and Communication Engineering,
Hainan University, Haikou, 570228, China and with the School
of Electronic and Optical Engineering, Nanjing University of Science and
Technology, Nanjing, 210094, China.}}

\thanks{\emph{
Chuan Ma is with Zhejiang Lab, Hangzhou, China, with School of Electrical and Optical Engineering, Nanjing University of Science and Technology, Nanjing, China and with Key Laboratory of Computer Network and Information Integration (Southeast University), Ministry of Education, China.}}

\thanks{\emph{
Jiangzhou Wang is with the School of Engineering, University
of Kent, Canterbury CT2 7NT, U.K.}}
}

\maketitle
\begin{abstract}
  Recently,  the  development of mobile edge computing has enabled exhilarating edge artificial intelligence (AI)  with fast response   and low communication cost. The location information of edge devices is  essential to support the  edge AI in many scenarios, like smart home, intelligent transportation systems and integrated health care.  Taking advantages of deep learning intelligence, the centralized machine learning (ML)-based positioning technique has received  heated attention from both academia and industry. However, some potential issues, such as  location information  leakage and  huge data traffic, limit its application.  Fortunately, a newly emerging privacy-preserving distributed ML mechanism, named federated learning (FL), is expected  to alleviate these concerns. In this article, we illustrate a framework of FL-based localization system as well as the involved entities at edge networks. Moreover, the advantages of such system are elaborated.  On practical implementation of it, we investigate the  field-specific issues associated with system-level solutions, which are further demonstrated over a real-word database.  Moreover, future challenging open problems in this field are outlined.
\end{abstract}

\section{Introduction}
Recently, the advance in  wireless communication technologies  and artificial intelligence (AI) has promoted orders-of-magnitude increases  of interconnected smart devices, i.e., internet of things (IoT). To provide smart context-aware services  with  massive  data generated at the network edge, mobile edge computing (MEC) has been developed rapidly \cite{DBLP:journals/iotj/ShiCZLX16}. By sufficiently exploiting the communication/computing resources at edge networks,  MEC enables ultra-low latency, high-bandwidth, and real-time edge AI.

The position information of  edge devices is a cornerstone to  enable sound and real-time  operations of the edge network. On  one hand, position information can be utilized across all layers of the communication protocol stacks to design and optimize the communication system at  edge networks \cite{DBLP:journals/spm/TarantoMRSSW14}.
On the other hand, position information is  naturally  indispensable on the application of location-based  services (LBS), such as navigation, target tracking, recommending system and mobile game. As the era of Beyond 5G (B5G) and IoT arrives, brand-new  LBS will emerge  in a wide range of application areas \cite{DBLP:journals/network/LiXWZWS19}, including industry 5.0, smart home, intelligent transportation systems, etc. For example, after acquiring self-positioning, the industrial robot is able to cooperate with  others in a highly automated way. Therefore, how to accurate predict the location is always a key component to support edge AI.

As the rapid development of  AI, machine learning (ML) technologies have played an important role in providing position information in  complex environments \cite{DBLP:journals/network/LiXWZWS19}. ML-based localization consists of two phases: the off-line phase and the on-line  phase. In the off-line  phase, a ML model is trained by learning the relationship between the location-dependent measurements and  related positions. In the on-line  phase, the trained model is used to predict the real-time  position of a device by requiring the location-dependent measurement of it.

Although ML-based localization has received extensive attentions, there are two critical issues that limit its practical applications:  privacy concerns and  huge data traffic. Traditional centralized learning based technologies require uploading raw data (measurements and related positions) of participated devices, named clients hereinafter,  to a central server, which generates huge data traffic  especially in large-scale systems.  Also, in this process, position information  is exposed directly, and may be  intercepted by an adversary.

Recently, a distributed ML mechanism, named federated learning (FL)  \cite{A17} has gathered tremendous interests. In FL, local ML models are trained at clients where training data is generated, and a global ML model is generated in the central server by  aggregating the local models. This cooperative leaning is completed by exchanging local model parameters  rather than the massive raw data that contains privacy information. With the potential of addressing the issues of  privacy concerns and  huge data traffic, FL-based localization has been researched, and has shown considerable prediction performance  \cite{9250516,9103044,9761235}.

In this paper, we  first illustrate the process of FL-based localization  at the edge network, and highlight the involved entities and their operations. Different from existing works \cite{9250516,9103044,9761235}, we also point out three inherent issues including measurement heterogeneity, environmental variation and 3D localization. This article aims to discuss these vital aspects in the field of FL-based localization and provide system-level solutions.
The remainder of this article is organized as follows.  In Section II,  the framework of  FL-based localization at the mobile edge network is illustrated and  its main advantages are elaborated. Then in Sections III and IV, we focus on three field-specific challenges when implementing FL-based localization, and provide potential solutions.  Subsequently, some key opening problems that deserve future researches are discussed in Section V.  Finally, conclusions are drawn in Section VI.

\section{FL-based Localization System at the Edge Network: Procedures and Advantages}
\begin{figure*}
  \centering
  \includegraphics[width=1\textwidth]{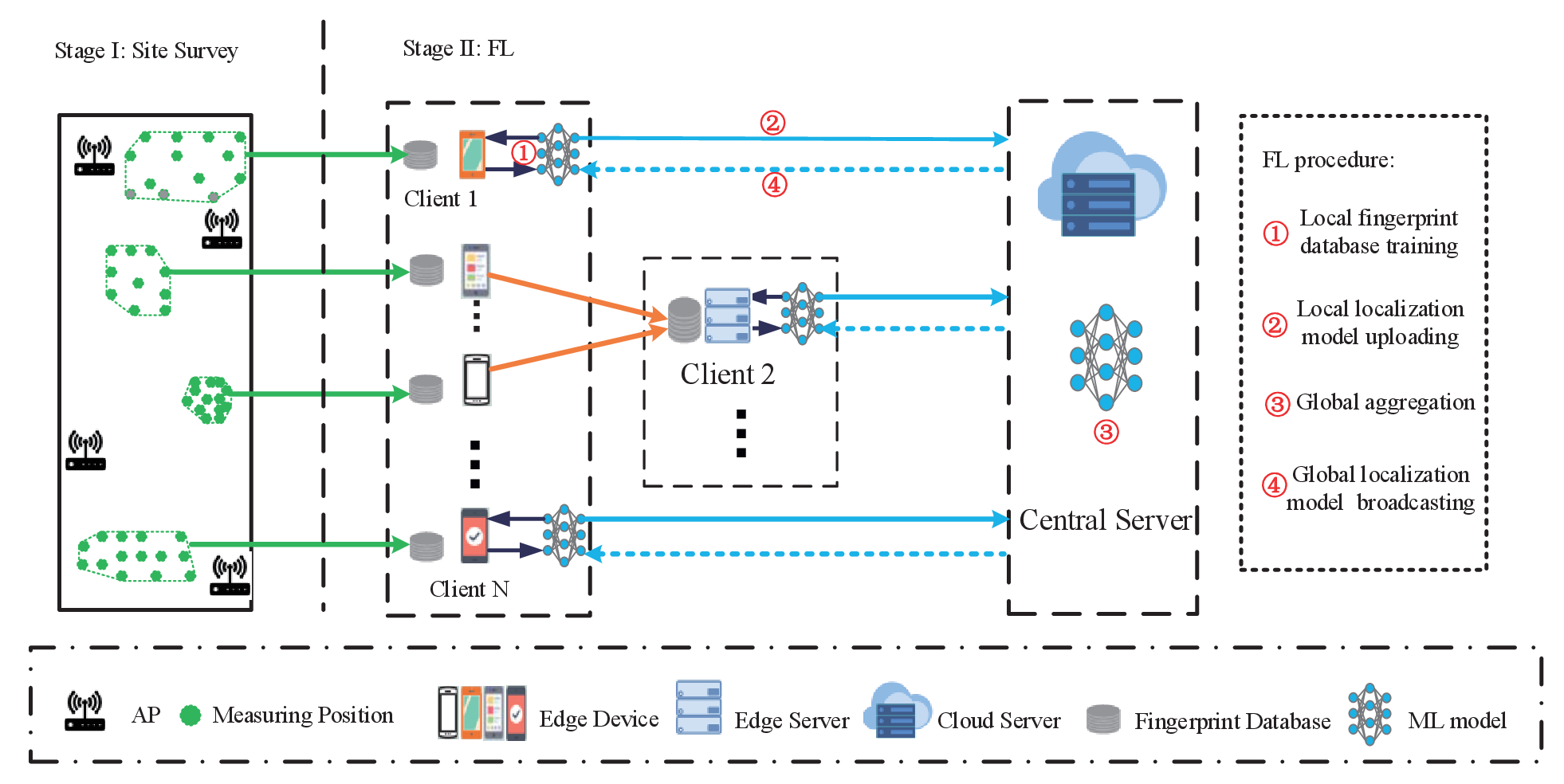}\\
  \caption{Schematic diagram of FL-based localization at the edge network.}\label{flow}
\end{figure*}

In this section, we first illustrate the framework of FL-based localization system at the edge network  by two steps as shown in Fig.~\ref{flow}, and then elaborate main advantages of the proposed framework.

\subsection{Stage One: Fingerprint  Database Construction}
The first stage of FL-based localization  is to construct training databases, usually termed as fingerprint database. Specifically, active edge devices perform  a site survey over the area of interest (AoI) assisted with access points (APs), as illustrated in Fig.~\ref{flow}.  APs are deployed at fixed positions and broadcast  signal  for location sensing. Ambulatory  edge devices pass a certain number of  positions in the AoI while reading the location-dependent patterns, such as received signal strength of  AP signals  at each position.  Finally, each edge device storages several data pairs as local fingerprint database, consisting of the measurements of AP signals (features) and corresponding measuring positions (labels). We next describe the properties of the  two involved entities.

\emph{APs}  refer to any fixed devices that are able to emit the signal used for location sensing, mainly including WiFi, Bluetooth,  ultra-wideband (UWB), radio-frequency identification(RFID), and ultrasound \cite{DBLP:journals/network/LiXWZWS19}. The type of AP signal  determines the advantages and disadvantages localization system directly. Among them, WiFi-based localization system  has developed most widely and achieved most successful application due to  low hardware cost \cite{DBLP:journals/network/LiXWZWS19}.

\emph{Edge devices} participating in the site survey are equipped with  logging,  storage, and sensing entities, which are  integrated in overwhelming    smart devices  such as smart phone, intelligent industrial robot, wearable device and  UAVs.

\subsection{Stage Two: Federated Learning Process}
Fingerprint database are constructed at edge devices in the site survey, and can be used to train a localization engine under the FL framework.    In this manner, local clients  collaborate in training a ML model under the coordination of a central  server with communication of model parameters.  A complete picture of FL structure at the  edge network is depicted in Fig.~\ref{flow}. The FL is an iterative process, and the procedures at each round of FL contains the following four steps.
\begin{itemize}
  \item \textit{Local fingerprint database training:} each client updates the local model  in parallel based on the global model parameters, received from the central server. Updates are completed by optimizing the local model parameters to minimize the training loss over a local fingerprint database.

  \item \textit{Local localization model uploading:} each  client transmits the  updated local model parameters to a central server.

  \item \textit{Global aggregation:} the central server aggregates the received local models according to the calculated weights, and   updates the global model.

  \item \textit{Global localization model  broadcasting:} the central server broadcasts the updated global model parameters to selected clients for the next round of learning.
\end{itemize}
The above steps are repeated  until convergence where the update of global model grounds to a halt. Finally, the last global model can be put into use in the on-line phase.  We next illustrate the  properties and operations of involved entities in FL at the edge network.

\emph{Clients}  train the local model over fingerprint database and upload the updated local model parameters to the central server.  Therefore clients are equipped with logging, computing, storage, and communication entities.  There are two types of clients in the edge network. The edge device that has done  the site survey  can  participate in the FL  as a local client. Moreover, trustful edge servers can become a local client as long as having access to the fingerprint databases of edge devices. To further protect the privacy of these fingerprint databases, we can utilize the differential privacy or encryption  based techniques to prevent the privacy leakage \cite{9347706}.  By undertaking the learning task from some communication/computation-limited edge devices,  the edge servers enable a reliable and real-time learning process .

\emph{The central server} refers to the cloud server,  equipped with  high-speed computing, cache/storage and communication entities. Cloud server aggregates the  local training model uploaded from distributed clients to update the global model and distributes it to all clients. Note that the aggregation is usually a low-complexity operation like averaging, so  the cloud server can synchronously handle other  tasks.

\subsection{Why  FL-based Localization?}
In what follows, we will elaborate the irresistible reasons of choosing the above FL-based localization system, compared to traditional localization technologies.

\emph{Data-based localization:}  The fingerprint databases collected in the site survey match the underlying mechanism of the complex  environment on the AoI. Learning the fingerprint databases in FL-based localization  rather than building the signal propagating model as in model-based localization is more suitable for complex environments  where many edge devices may be located in.

\emph{Large-scale application:} In the online phase, any edge device in the AoI can infer its position after receiving  the converged global model from the central server. Note that the devices are not limited to FL clients.  Moreover, the neuronic network  model or Gaussian process model  are two representative inference  engines in FL-based localization \cite{9250516}. The on-line location inference of such engines usually takes extremely short response and  low computing resource. It can be concluded that the universal, low-latency and low-complexity location inference enables the large-scale application of such engines at the edge network.

\emph{Privacy-preserving and communication-efficient mechanism:} In FL, the cooperation of distributed clients is achieved through communication with a central server, and only local localization model parameters  instead of raw data are transmitted. This mechanism can save extra communication resources for both clients and the central server, compared to the centralized ML. Note that communication resource is crucial to provide better quality of service (QoS)  for users in edge networks. Moreover, such cooperative mechanism greatly avoids the leakage of  location information  of clients from its fingerprint database to the external third-party. It is  essential to promote the application of location-aware services.  Because the highly sensitive location information  and the personal behavior reflected by it can be utilized by adversaries or eavesdroppers, causing potential troubles and risks \cite{9103044}.

\emph{Liberation of the central server:} In FL, the central server acts as an assistor that helps aggregate the local  models trained by local clients, and does not need to perform the complex model training task. Therefore, only a small share of computing resources in the central server  is occupied by FL-based localization, and the central server can synchronously handle other vital tasks at the edge network.

\emph{Coordination of unbalanced resource:} The distributed nature of FL  ensures system resilience and service continuity of edge networks where unbalanced computing and communication resource is witnessed. In FL, each client learns its local model  in parallel at each epoch, and it is possible to optimize the FL process under the heterogeneous resources constraints  with advanced techniques \cite{DBLP:journals/cm/KhanPTSHNH20}.

\section{FL-based Localization Across Different Domains}
In this section, we first discuss two important and practical issues for implementing the FL-based localization, and then provide a system-level solution.

\subsection{Issues Description}
\emph{Measurement heterogeneity}: In the site survey,  different types of edge devices  are used to detect and measure  AP signals. Their different hardware facilities such as build-in sensors result in  inconsistencies of the detection and measurement  at the same position and  time. Therefore, the fingerprint databases collected by different type of devices  may  follow  non-independent identically distribution (non-iid). Note that there are many factors causing the measurement heterogeneity between clients, such as temperature and humidity, in this issue, we only focus on different types of devices.

\emph{Challenges:} FL aims to generate a generalized model for  balancing  the requirement of each client. Due to the severe measurement heterogeneity,  the FL model may be unsatisfactory  when applied on the target type of device. A common idea is to  train an  extra  type-specific FL model under the cooperation of target edge devices.  However, the fingerprint data collected by a single type of devices may be insufficient, so that training a model over the insufficient dataset may cause the over-fitting problem.

\emph{Environmental variation:} Another practical factor is the substantial environment variation in the AoI in different time phases. Signal propagation environment in the AoI always changes over the time, caused by  unpredictable activities of humanity,  movements of objects, and even the variation of temperature and humidity. Besides, some APs  may be shifted to different locations. All these factors result in the  distribution discrepancy of   fingerprint measurements over the AoI between different time phases.

\emph{Challenges:} The existing FL model is trained by a large number of data collected at the time phase of  the site survey, and thus it only specializes in learning the feature  during that time phase. Applying the existing FL model directly to predict locations  in different time phases may be unsatisfactory. A common idea is to retrain an  extra time-specific FL  model over new fingerprint databases, which are collected  at the target time phase. However, re-collecting massive fingerprint data is highly-cost and time-consuming.  Therefore, the  newly collected fingerprint data is usually insufficient to retrain the time-specific FL  model.

\subsection{A Federated Transfer Learning-Based Approach}
In essence, the mentioned two issues can be integrated  as the statistical heterogeneity between training domain and application domain. In the issue of measurement heterogeneity,  the training domain refers to the mixed fingerprint data distribution on diverse edge devices, while the application domain refers to that on the target type of devices. In the issue of environmental variation,  the training domain means the fingerprint data distribution during the time phase of the site survey,  while the application domain means that at the target time phase.

Transfer learning (TL), which focuses on  transferring the knowledge learned from the source domain  to  different but related target domain \cite{5288526}, is greatly suitable  for this scenario.  Therefore, utilizing  TL in the  FL-based localization, i.e., federated transfer learning (FTL)-based localization,  seems to be a great solution. FTL  has been developed recently and successfully  applied to provide personalized AI in multiple regions, including smart hearth, human mobility prediction, and so on \cite{9090366}.

Then, we propose a hybrid federated transfer learning-based localization scheme (H-FedTLoc), which enjoys general knowledge sharing from traditional FL and meanwhile specific  knowledge owned from target-specific training. The key idea is to transfer the global FL model, trained by large-scale data on source domain to  sub-global FL models, which will be fine-tuned over  small-scale trainable data on target domains. Different from existing FTLs that focus on local personalization \cite{9090366}, the proposed H-FedTLoc focuses  on local-global personalization using a two-layers FL framework. The details are illustrated as follows.

As shown in Fig 2, the whole process consists of the following three steps:
\begin{itemize}
  \item \textit{Global FL:} each participant collaborates in training a high-quality global  model with the assistance of the central server in a  FL manner.

  \item \textit{Model transfer:} after global FL, the central server transfers the  global model to a sub-global model built for  further personalization on target type of devices.

  \item \textit{Sub-global FL:} the central server transmits the sub-global model  to clients, which  have access to the target database (fingerprint database collected by target type of devices). Starting from the received models,  the related clients collaborate in training the sub-global  model over local target databases in a FL manner.
\end{itemize}

As for the issue of environmental variation, the  basic process of H-FedTLoc follows the three steps, but the target database becomes the newly collected database  at the target time phase. Note that by transferring some knowledge from the source domain, the H-FedTLoc may have a faster training speed than the federated training over the newly collected data. Therefore, the H-FedTLoc is expected to be effective in time-sensitive localization tasks.

\begin{figure}
  \centering
  \includegraphics[width=0.45\textwidth]{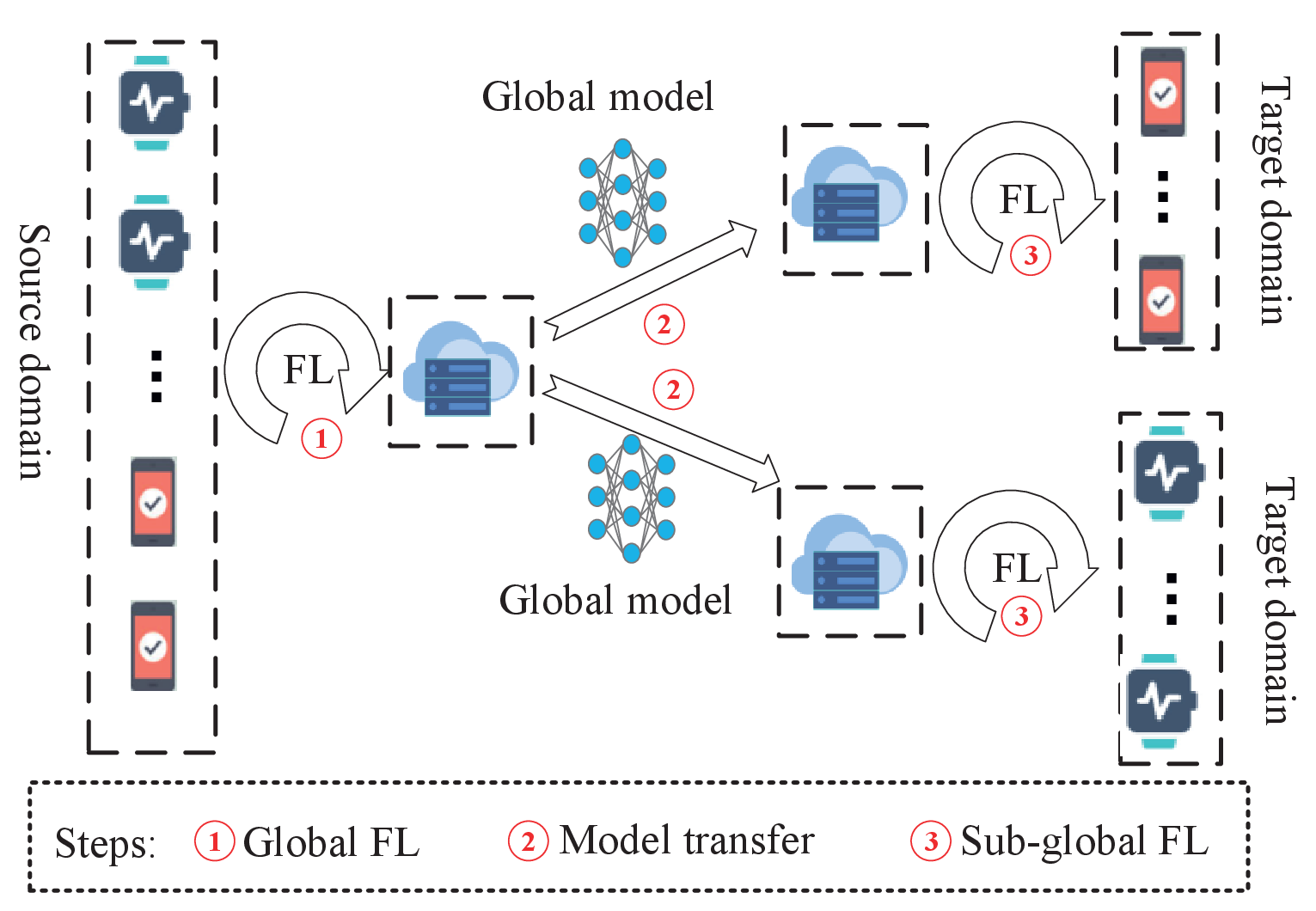}\\
  \caption{Schematic diagram of  H-FedTLoc across devices.}\label{F_HFedTLoc}
\end{figure}

\section{FL-based Localization in 3D Cases}

\subsection{FL-based 3D Localization}
Positioning in a three-dimensional (3D) case, i.e, a multi-floor building, has  attracted extensive attentions.  However, a direct predicting of  the three dimensional position of a edge device in a multi-floor building  usually has poor precision, and an effective approach is to predict in two steps \cite{7275492}. In the first step, a ML-based classifier is used to determine the height of the location, e.g., which floor the device locates on.  In the second stage,  the two-dimensional (2D) position of the device on the located floor is predicted. Based on this, an  extension of the FL-based localization to 3D cases is proposed, which is straightforward and handy to implement.

In the off-line phase, to construct local  fingerprint databases, a  site survey is conducted in the multi-floor building by clients. After the site survey, a floor classifier is first learned with the coordination of  all local clients  in  this building under the  FL framework. Then for each floor, a 2D localization model is learned by local clients on the floor.  In the on-line phase, using the floor classifier and  floor-specific ML  models, the position of device can be predicted quickly. Therefore, FedLoc-3D consists of a FL-based floor classifier and 2D localization models.

\subsection{Heterogeneous Scenario}
Different from the centralized 3D localization, FL-based 3D localization suffers from  extreme unbalanced data distribution between local databases. It is non-trivial for some clients, such as  robots to do cross-floor site survey. For a client such as a smart phone carried by a person, performing cross-floor site survey is time-consuming and labor-costing. Therefore, usually,  a client  is only active on a single floor in the building during the site survey.  In the multi-class floor classify task, each local database may only consists of one-class data, causing serious statistical heterogeneity. To this end, a simple but elegant FL-based  training scheme, namely  federated one-vs-all (FedOVA) \cite{9609654}, is introduced to predict the located floor of devices.

\begin{figure}
  \centering
  \includegraphics[width=0.45\textwidth]{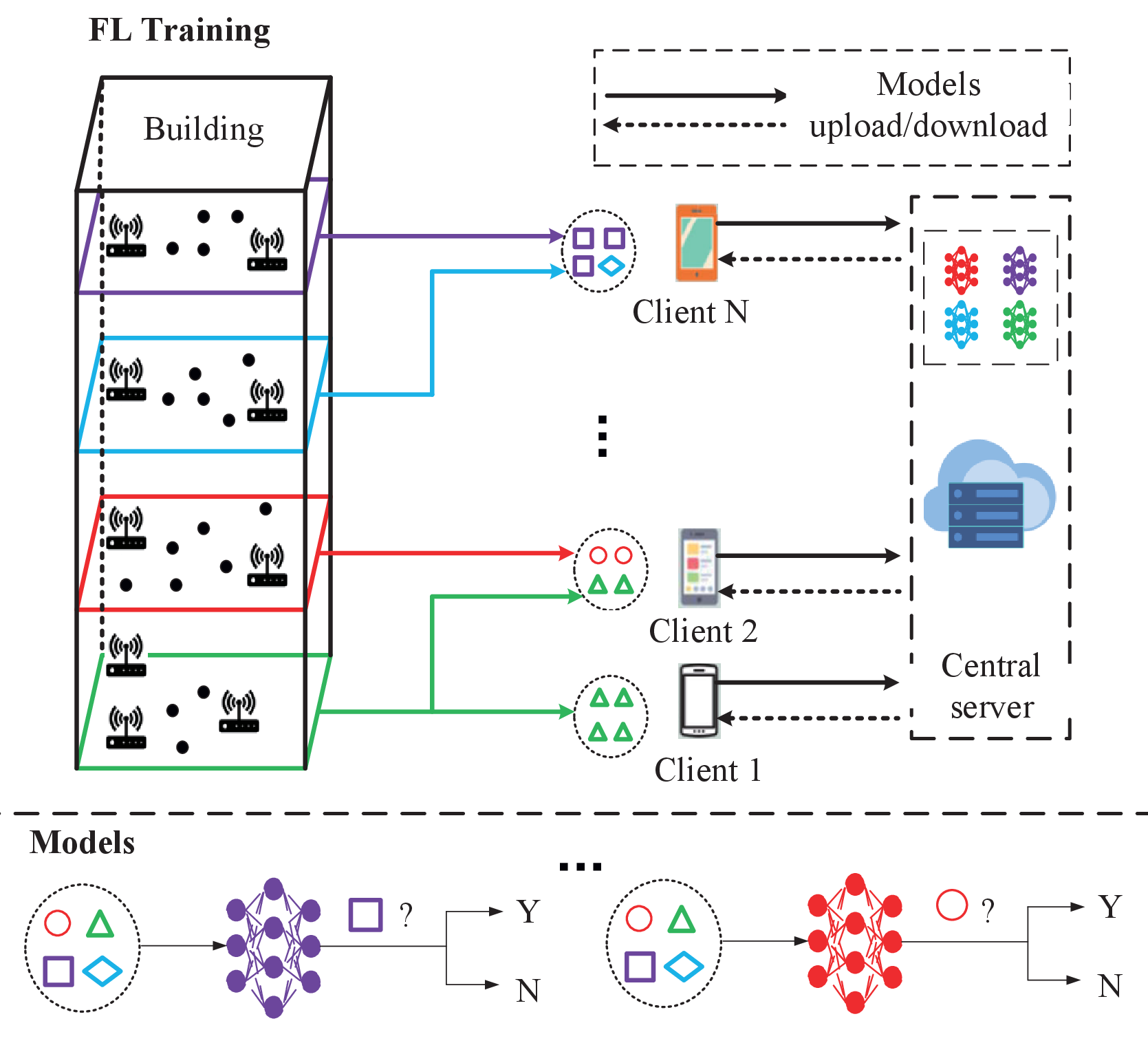}\\
  \caption{Schematic diagram of  FedOVA-based floor classifier in a 4-floor building.}\label{F_FedOVA}
\end{figure}

FedOVA aims to decompose a multi-classification task into multiple binary classification tasks under the FL framework. The  procedure of FedOVA-based floor classifier is illustrated in Fig.~\ref{F_FedOVA}.  Specifically, for a $L$-floor building, totally $L$ binary classifiers are learned,  each of which specializes on a single but different floor. Such classifier aims to output the probability of locating on the focused floor or not.  In the off-line phase, totally $L$ models are trained independently. Note that, each client takes local update  on  the whole $L$ binary classifiers. In the aggregation step, the central server aggregates the floor-specific models separately.  In the on-line phase, a device can input its real-time fingerprint measurement to the trained $L$  classifiers, and then obtain the probabilities of locating on each floor in the building. Naturally, the floor with the max probability is selected as the prediction result.

\section{Experimental Results}
In this section, we provide experimental results on a widely utilized real-word database, named UJIIndoorLoc \cite{7275492} to demonstrate the aforementioned issues and associated solutions.


\emph{Database description}:  UJIIndoorLoc is collected at three buildings with four or more floors of the Jaume I University (UJL) that covers almost 110000  $m^2$. It provides 21049  sampled points where 19938 is for training and 1111 is for validation, which is conducted by more than 20 people and 25 different devices during several months. The fingerprint measurement in this database is  the received signal power (RSS) of WiFi signals.

\emph{Basic settings}: The 4-floor build with identification equal to 1 is selected for 3D cases and the floor with identification equal to 1 in this building is selected for 2D cases. For 2D cases, the client amount is set to be 8 while for 3D cases, the client amount is set to be 16. TensorFlow libraries are utilized to implement the learning process by a MLP network. For model training, the initial and also the most common  FL scheme, i.e., Federated Averaging (FedAvg) \cite{A17} is used here.

\begin{figure}
	\centering
	\subfloat[On the target type of  devices.]{\label{fdevice}\includegraphics[width=0.45\textwidth]{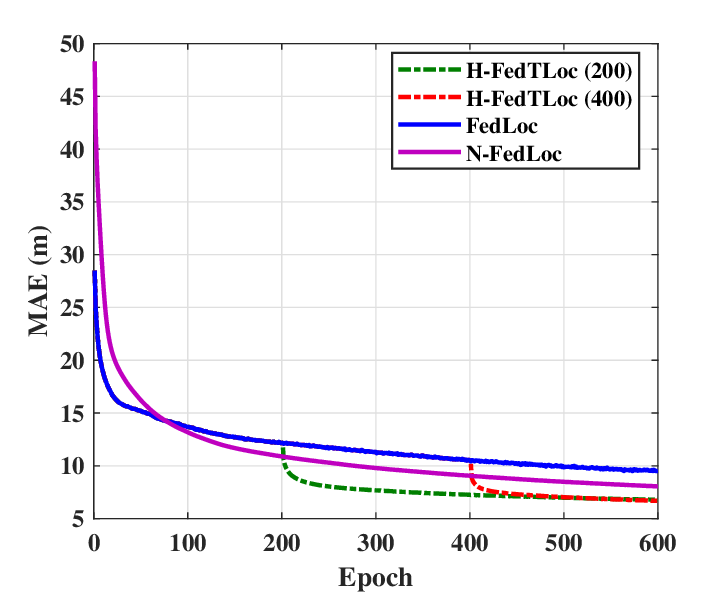}} \\
	\subfloat[At the target  time phase.]{\label{ftime}\includegraphics[width=0.45\textwidth]{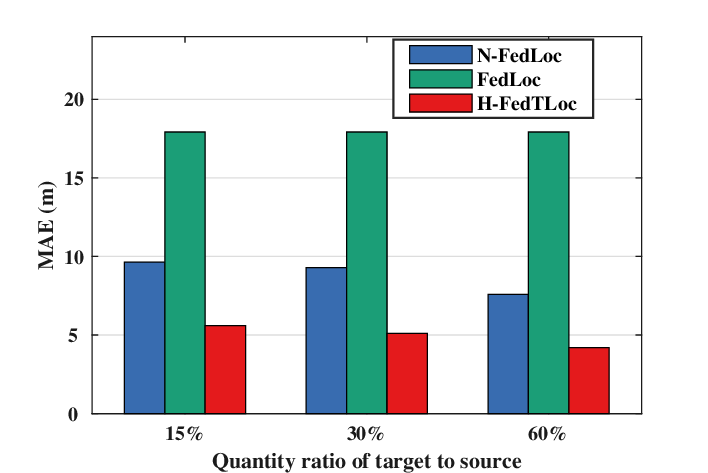}}
	\caption{Prediction error on the target domain.}
    \label{ftransfer}
\end{figure}

\subsection{Measurement Heterogeneity}
The discussed issues of measurement heterogeneity caused by different types of devices are well verified in the UJIIndoorLoc, since RSS values measured by different type of smart phones are obviously inconsistent  at the same position and time.  Note that in this scenario, each client only owns one device.  Three focused methods including FedLoc, N-FedLoc and our proposed  H-FedTLoc, are investigated  in this subsection.  FedLoc means training a FL model with all types of devices  without transfer,  and  N-FedLoc represents  newly training a FL model over trainable data on target type of devices.

Test MAE on the target type of device (one type of device that participate in the federated training) via training epochs  is provided in Fig.~\ref{ftransfer}\subref{fdevice}.  As seen, after transferring the global model  to the sub-global model in H-FedTLoc, the  prediction error can be reduced significantly within few rounds of epoches. Besides, it may be advantageous to perform the model transfer at an earlier epoch for enjoying a faster performance improvement.  At the final round (400 epochs),  the proposed H-FedTLoc reduces the prediction error substantially  compared to  FedLoc   and achieves an  almost 20\% performance gain over  N-FedLoc.

\subsection{Environmental Variation}
The discussed issues of environmental variation is well verified in the UJIIndoorLoc since the distribution of RSS values measured at different time phase are obviously inconsistent. The  interval between the source time phase  and the target time phase is about one month in this setting.
The aforementioned three  methods are also compared in this subsection. Differently,  FedLoc means training a FL model over data collected at the source time phase (source trainable data),  and  N-FedLoc  represents  training a FL model over trainable data collected at the target time phase (target trainable data).

Fig.~\ref{ftransfer}\subref{ftime} illustrates test MAE at target time phase via training epoches with various quantity ratios  of target trainable data  to source trainable data. This result indicates that  the performances of H-FedTLoc and N-FedLoc improve as the amount of target trainable data increases. Moreover, in both scenarios, the performance of the proposed H-FedTLoc surpasses  FedLoc obviously and reduces the prediction error substantially  compared to  N-FedLoc.

\subsection{3D Cases}
\begin{figure}
  \centering
  \includegraphics[width=0.45\textwidth]{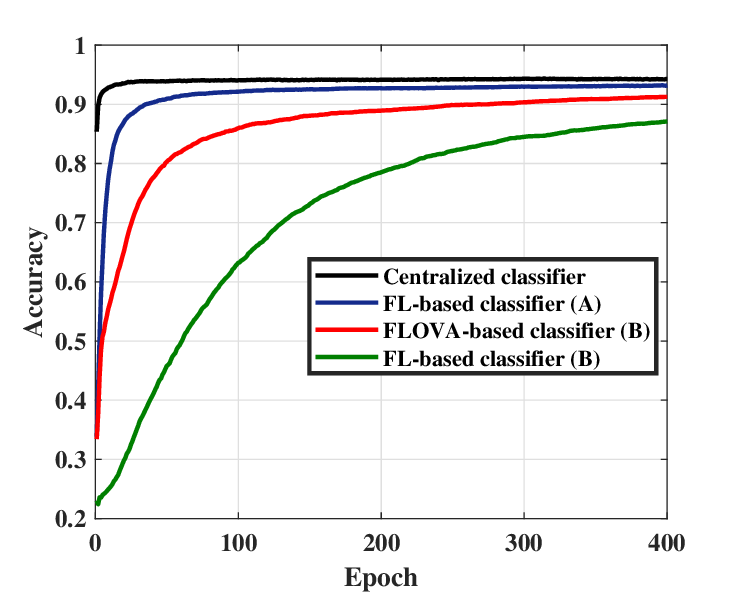}\\
  \caption{Test accuracy versus global  epoches.}\label{FL_floor_tu}
\end{figure}

In this subsection, we only focus on the FL-based classifiers  in the 3D case. After predicting the floor, the localization task becomes a  2D one, which has already been discussed.  We consider two scenarios. In the homogeneous scenario (scenario A), each client collects local database across the floors in the building. In  the heterogeneous scenario (scenario B), each client only collects local database on a single floor.  Fig.~\ref{FL_floor_tu} shows the testing accuracy  versus global training epoches. As can be observed,  FL-based classifier  converges fast and has a near-centralized accuracy in  the scenario A but performs poorly in  the scenario B.  In the scenario B, the test accuracy of the FedOVA-based classifier  surpasses  the FL-based classifier obviously at the epoch of 400.

\section{Challenges and Open Research Topics}

\subsection{Label-less FL-based localization via Crowdsourcing}
Recently, fingerprint crowdsourcing  has been extensively studied, which  relieves the burden of site survey from  professional surveyors to common users such as mobile devices in a participatory sensing manner \cite{8529230}.  But there arises an important issue that is common users may  be unwilling or unable to explicitly label fingerprint data with the location information as the professional surveyors do in the site survey. Therefore,  most or whole fingerprint data is unlabeled in localization via crowdsourcing.

A common idea to address this issue by performing location annotation with little user intervention. In this way, several systems have been designed by using the inertial sensors of devices with the aid of a floor plan or without \cite{8529230}.  However, existing schemes are centralized, labeling isolated local fingerprint databases with little user intervention for FL-based localization via crowdsourcing deserves further works.

In \cite{9103044}, authors propose a centralized indoor localization method using pseudo-label (CRNP) which leverages the power of unlabeled fingerprint data, and then incorporate CRNP and FL-based localization system. Inspired by this, exploiting more effective semi-supervised/unsupervised learning in FL-based localization system can be investigated as future directions.

\subsection{Heterogeneous Spatial Distribution}
On the practical implementation of FL-based localization,  unbalanced client behaviours emerges due to the differences of built-in hardware and  located areas on the AoI. Such unbalanced behaviours  including  sampling intervals, sampling amount and trajectories of each client will result in  the spatial heterogeneity  of local databases \cite{9761235}. Such spatial heterogeneity leads to unbalanced quality between local databases. Generally, it is more reasonable to distribute larger aggregation weights to local model trained over higher-quality fingerprint databases rather than averaging them in FedAvg \cite{A17}.

Authors in \cite{9761235} characterized the local  database quality  by the area of sampling convex hull and showed the effectiveness of distributing the aggregation weights according to the characterized qualities directly. However, this method only focuses on the sizes of local sampling areas  but ignores the relative spatial relationship  of these areas.  Designing more comprehensive aggregation weights from both theory and practice deserves further works.

\subsection{Non-IID cases in FL-based Localization}
The non-iid cases are practical and inevitable in the FL-based localization. In this work, we consider three forms of non-iid data scenarios in localization, measurement heterogeneity, environmental variation and heterogeneous 3D localization. In addition,  heterogeneous spatial distribution \cite{9761235} is also a non-iid data scenario. Besides the technologies we introduced, many techniques in the FL framework are responsible for handling the non-iid data, including  meta learning, multi-task learning and knowledge distillation \cite{9090366}. Using these technologies   to solve the inherent issues in FL-based localization and compare their performance deserve further researches.

\subsection{Time-sensitive FL-based Localization}
Getting location information with fast response is crucial to some time-sensitive servers such as  intelligent transportation systems. Reducing the time delay of FL-based localization is  necessary since the FL may convergent slowly with non-iid data or unbalanced resources.
Many technologies are researched to reduce the time delay of distributed ML, including designing gradient descent optimizer and coordinating/optimizing the computing/communication resources \cite{9534784}. Absorbing these technologies in FL-based localization methods is expected to provide location information with fast responses as a further direction.

\section{Conclusion}
In this article, the framework of FL-based  localization  at the edge network as well as its advantages has been illustrated. On the practical implementation, we have illustrated three inherent issues including  measurement heterogeneity, environmental variation and  3D localization as well as provided system-level solutions.  The effectiveness of these system-level solutions have been demonstrated on the real-word UJIIndoorLoc database.  Finally, we have  outlined other challenging problems which deserve further researches.

\ifCLASSOPTIONcaptionsoff
  \newpage
\fi
\small
\bibliography{IEEEfull,cite}
\bibliographystyle{IEEEtran}

\end{document}